# Strain-modulated Intercalated Phases of Pb Monolayer with Dual Periodicity in SiC(0001)-Graphene Interface

Lin-Lin Wang[1*], Shen Chen[1,2], Marek Kolmer[1], Yong Han[1,2] and Michael C. Tringides[1,2#]

[1]Ames National Laboratory, U.S. Department of Energy, Ames, IA 50011, USA
[2]Department of Physics and Astronomy, Iowa State University, Ames, IA 50011, USA

*llw@ameslab.gov
#mctringi@iastate.edu

## Abstract

Intercalation of metal atoms at the SiC(0001)-graphene (Gr) interface can provide confined 2D metal layers with interesting properties. The intercalated Pb monolayer (ML) has shown the coexistence of the Gr(10×10)-moiré and a stripe phase, which still lacks understanding. Using density functional theory calculation and thermal annealing with ab initio molecular dynamics, we have studied the formation energy of SiC(0001)/Pb/Gr for different coverages of intercalated Pb. Near the coverage of a Pb(111)-like ML mimicking the (10×10)-moiré, we find a slightly more stable stripe structure, where one half of the structure has compressive strain with Pb occupying the Si-top sites and the other half has tensile strain with Pb off the Si-top sites. This stripe structure along the Gr zigzag direction has a periodicity of 2.3 nm across the [1$\bar{2}$10] direction agreeing with the previous observations using scanning tunneling microscopy. Analysis with electron density difference and density of states show the tensile region has a more metallic character than the compressive region, while both are dominated by the charge transfer from the Pb ML to SiC(0001). The small energy difference between the stripe and Pb(111)-like structures means the two phases are almost degenerate and can coexist, which explains the experimental observations.

# I. Introduction

Moiré structures of few-layer graphene (Gr) systems[1-3] with twisting have shown many interesting behaviors by tuning electron interaction and correlation. Although metal atom intercalation in the epitaxial Gr grown on SiC(0001)[4] was first aimed to decouple Gr away from the SiC substrate, it also provides a promising approach to realize ultrathin 2D metal layers in confined environments[5] that coupled to the 2D Dirac electron of Gr. This platform offers opportunities for future applications by tuning Gr band structure depending on the metal. Among them, Pb intercalation in epitaxial Gr on SiC(0001) has attracted intensive studies [6-19], including the moiré pattern with the 10-fold periodicity of Gr(10×10) [6, 12, 17]. Besides the (10×10)-moiré, the coexistence of a stripe phase has also been observed in several scanning tunneling microscopy (STM) studies[6, 8, 9, 13]. The structures of the intercalated Pb in SiC(0001)/Gr have been modeled from a Pb(111)-like full monolayer (ML) occupying all the dangling Si-top sites[14, 15] also possibly with domain boundaries[16] to a (7×7) Pb(111)-like layer fitting the Gr(10×10) supercell[8, 12], while the stripe phase has been proposed as two Pb(110)-like layers[8]. Recent density functional theory (DFT) studies have explored a range of SiC(0001) supercells with increasing sizes and various Pb coverages[11, 17-19]. However, the key question of why a stripe phase and the (10×10)-moiré for SiC(0001)/Pb/Gr can coexist is still not answered in terms of relative thermodynamic stability. Especially with the proposed one-layer Pb(111) for the (10×10)-moiré and two-layer Pb(110) for the stripe phase[8], the domain out-of-plane atom distribution and Pb coverage of such two structures are quite different.

Here using DFT calculation and thermal annealing[20, 21] with ab initio molecular dynamics (MD), we study the formation energy of SiC(0001)/Pb/Gr as a function of Pb intercalation coverage. We find a Pb ML stripe structure is as stable as the Pb(111)-like structure around the same ML coverage to explain the coexistence of the (10×10)-moiré and stripe phases. Given that both previous experiments[6, 7, 10] and DFT studies[11, 17-19] have found the most stable location of Pb intercalation is beneath the Gr buffer layer (BL) and there is a good alignment between the single-layer Gr (SLG) and the decoupled BL from SiC(0001) after Pb intercalation, we focus on the Pb ML under the SiC(0001)/Gr BL in this study with the $(5\sqrt{3} \times 5\sqrt{3})$ SiC(0001) supercell supporting the Gr(10×10) supercell. We find the highest ML coverage with all the Si-top site being occupied has a large compressive strain. After thermally annealing with MD for more stable structures at lower coverage, we discover a strain-modulated Pb ML which shows a stripe pattern along the main in-plane axis direction being 8 meV/Pb more stable than the Pb(111)-like structure at almost the same coverage, thus explain the coexistence of a stripe phase and the (10×10)-moiré for SiC(0001)/Pb/Gr. The DFT-calculated intercalated ML Pb stripe structure is along the Gr zigzag direction and has the periodicity of 2.3 nm across the $[1\bar{2}10]$ direction, which agrees well with the direction and periodicity of the stripe phase found in STM experiments[6, 8, 9, 13]. In this stripe structure, one half of the structure has the compressive strain with Pb occupying the Si-top sites and the other half has the tensile strain with Pb off the Si-top sites. Our analysis on electron density difference and density of states show that there is competition between the need to passivate Si dangling bonds and maximize Pb-Pb interaction at the SiC(0001)/Gr interface, which results in strain modulation and dual periodicity. We find the tensile region has a more metallic character than the compressive region, while both are dominated by the charge transfer from the Pb ML to SiC(0001). Thus, different structures with energy degeneracy compete and coexist, which can be affected and tuned by different growth and annealing conditions of Pb intercalation.

## II. Results

### II-a. Structural types and phase stability

In experiments, Pb surpisingly intercalates at relatively low temperatures ~450 ºC as shown in several recent publications[7-9, 12]. This observation is unexpected given the large size of the Pb atom. Spot profile analysis low energy electron diffraction (SPA-LEED) is ideal to determine the intercalation of a given metal, especially the sub-surface location where it bonds to. The intensity drop of a spot originating from a sub-surface location measures the degree of intercalation. The Pb(10) spot, observed after deposition indicates the formation of Pb(111)-like islands of multiple height. When the Pb(10) spot becomes extinct it indicates island decay, release of Pb and onset of intercalation. The formation of the (10×10)-moiré depends on the initial surface condition with single layer graphene (SLG) and results in weaker spots[12, 16, 17] than the case of buffer layer. Figure 1 shows the 1D SPA-LEED profiles after stepwise deposition at LN$_2$ and annealing cycles of total 13 ML on a surface already intercalated by Pb. Gradual annealing from 200 ºC to 450 ºC was followed for 1h at each temperature. The diffraction profiles obtained on such pre-intercalated surface are along $[1\bar{2}10]_{SiC}$ (graphene reprocipical lattice direction). A new weak spot was observed at approximately 12.5%BZ$_{SiC}$ which corresponds to a Gr(10×10) supercell. Deposition and annealing steps were repeated for energies within 34-100 eV to be sensitive to the top layers. However, the new (10×10) spot was observed only at 44 eV (black) and 70 eV (magenta). Few energies to oberve the (10×10) spot were reported in other similar intercalation studies of Pb under SiC(000)/Gr[9, 12]. From the STM measurement in real space, the (10×10) phase also tends to appear together with small domains of stripe phase as shown in Fig.2 of Yurtsever *et al*[6], Fig.4. of Hu *et al*[8] and Fig.3 of Gruschwitz *et al*[9]. The observed stripe domains are very small as seen by STM and no diffraction spots have been seen originating from this phase, which is consistent with the small size of these domains.

Computationally, to model the Gr(10×10) and related phases for intercalating Pb in SiC(0001)/Gr, we consider and search for the most thermodynamically stable structures at different Pb coverages. Figure 2 presents the Pb-intercalated structures inside the SiC(0001)/Gr BL interface at the high-symmetry sites with three different coverages ($\theta_{gr}$) of 0.25 in (a-b), 0.50 in (c-d) and 0.75 in (e-f). The coverage, $\theta_{gr}$, is defined with respect to Gr primitive unit cell and the formation energy ($E_f$) is defined with respect to the SiC(0001)/Gr BL without any Pb and the bulk face-centered cubic (FCC) Pb (see Eqn.(1) in Methods). The green cell in Fig.2(a), (c) and (d) is the smallest supercell that can accommodate these structures in the Gr(2×2) supported on top of the SiC(0001) ($\sqrt{3} \times \sqrt{3}$) unit cell. But to compare with other structures that fit into the Gr(10×10) supercell, these structures have been calculated in the SiC(0001) ($5\sqrt{3} \times 5\sqrt{3}$) supercell. Because we are mostly interested in the (111)-like Pb ML structure and the highest coverage in Fig.2(e-f) with the in-plane Pb orientation being rotated 30 degree from that of Gr, we name these structures with the keyword of R30 then followed by the number of sites being occupied by Pb, such as R30-Si-3top, which means all the three Si-top sites in the green cell are occupied by Pb at the highest ML coverage. Such coverage corresponds to a high level of stress when compared to the lower coverage to be discussed below. Note this structure at the highest ML coverage was deduced from a different experiment with small density of defects and after annealing the deposited Pb up to 945 °C[14].

At $\theta_{gr}$=0.25 with one Pb in the ($\sqrt{3} \times \sqrt{3}$) unit cell of SiC(0001) in Fig.2(a-b), the most stable site for Pb is the hollow site just above a C atom in the second layer (1holC) instead of the Si-top site. The calculated $E_f$=+0.421 eV and site preference agree with the earlier studies[11, 19]. With increasing $\theta_{gr}$ to 0.50 for two Pb atoms in Fig.2(c-d), the Pb atoms themselves form a honeycomb, or a plumbene-like structure. This coverage is close to that of the (7×7) Pb(111)-like layer matching Gr(10×10) or Pb(111) (9×9) matching Gr(13×13) supercells. But due to the need to fit Gr(10×10) on top of SiC(0001) ($5\sqrt{3} \times 5\sqrt{3}$) supercell in computation, the Gr lattice has to be expanded by +8.4%. Thus the Pb(111)-like layer increases from (7×7) to (8×8) in our computational setup, which will be discussed later in more details. With Pb atoms directly interact with the two Si dangling bonds at the two Si-top sites, the SiC(0001)-Pb interaction is stronger than the smaller $\theta_{gr}$, but $E_f$=+0.151 eV is still positive, which means it is not thermodynamically stable with respect to SiC(0001)/Gr BL and bulk Pb. Moving to the highest coverage of a full Pb ML at $\theta_{gr}$=0.75, where all the three Si-top sites are occupied. The negative $E_f$ of –0.054 eV means it is now thermodynamically stable. This full ML Pb structure at high coverage has also been previously studied[14, 15]. But from the sideview in Fig.2(f), there is buckling of the Pb ML, which shows there is a significant compressive strain due to the high density of Pb at $\theta_{gr}$=0.75, higher than that of Pb(111).

The $E_f$ vs $\theta_{gr}$ are plotted in Fig.3 for the structures from Fig.2 together with the two other types of structures. For the high-symmetry structures from Fig.2, there is almost a linear decrease of $E_f$ (more stable) with increasing $\theta_{gr}$. It only becomes negative near the high $\theta_{gr}$ around 0.7, which shows that putting Pb directly on top of the Si dangling bond although maximize the interaction between intercalated Pb and SiC, it does not optimize the Pb-Pb interaction. The two other types of intercalated Pb ML structures can give negative $E_f$ already from $\theta_{gr}$=0.5. Later on, using thermal annealing with MD, we will show that at a lower coverage than $\theta_{gr}$=0.75, the strain can be relieved to create a stripe Pb ML structure as observed in STM.

For the second type of intercalated Pb structures, we construct a Pb(111)-(n×n) ML as the initial configurations for full relaxation, not requiring Si-top sites be specifically occupied. As shown in Fig.4(a-b), for (7×7) at $\theta_{gr}$=0.49 near 0.50, the fully relaxed structure with $E_f$ of –0.031 eV is much more stable than the $E_f$ of +0.151 eV for the R30-Si-2top structure of plumbene in Fig.2(c-d). The fully relaxed (7×7) structure is also more amorphous than the original (7×7) for optimizing the Pb-Pb interaction. It has more distorted hexagons than the honeycomb in plumbene. With $\theta_{gr}$ being increased to 0.64 for the (8×8) in Fig.4(c-d), the intercalated Pb(111) ML structure is stable upon full relaxation and maintains the original (8×8) lattice having small distortion with lateral separation between two Pb atoms closer to that in metallic Pb(111). Its $E_f$ of –0.144 eV is also much lower than the –0.054 eV for the highly-strained full Pb ML in the R30-Si-3top structure in Fig.2(e-f). Figure 5 plots the projected density of states (PDOS) per Pb for both the relaxed amorphous (7×7) structure in Fig.4(a-b) and that of the plumbene structure in Fig.2(c-d). Compared to the plumbene, the amorphous Pb PDOS is more spread out due to the more nearest neighbors (NN) than the three NN in the plumbene structure. The lower edge of the Pb 6$p$ orbitals around –2 eV is also shifted to lower energy for stronger bonding and higher stability.

## II-b. Stripe structure from annealing

The third type of intercalated Pb ML structures are the annealed R30-Si-top structures using MD. First at $\theta_{gr}$=0.49, to compare with the (7×7) structure, we removed one Pb from R30-Si-2top and annealed with MD until no lower-energy structure is found. Such annealed structure R30-Si-2top-m1-A is shown in Fig.6(a-b). Pb atoms form an amorphous structure, similar to that in the fully relaxed (7×7) in Fig.4(a-b), but the $E_f$ of –0.054 eV is lower than the –0.031 eV of (7×7). Then at $\theta_{gr}$=0.64, to compare with the (8×8) structure, we removed 11 Pb atoms randomly from the R30-Si-3top in Fig.2(e-f) and annealed with MD until no lower-energy structure is found. As shown in Fig.6(c-d), this structure has an interesting pattern despite being annealed in MD without any symmetry constraint. In the bottom half of Fig.6(c), Pb atoms are bonded on top of Si, except for the two Si-top sites in the green circle sharing only one Pb. This Pb atom also has a shorter distance from the Si plane due to sitting on the Si-bridge site as seen in the sideview of Fig.6(d). In the top half of Fig.6(c), many of the Si-top sites are not covered, but the Pb atoms still arrange in an almost regular pattern like an extended region from the bottom half with a tensile strain applied along the direction perpendicular to the *a*-axis, or the [1$\bar{2}$10] direction. These two patterns in the bottom and top half become clearer in Fig.6(e-f), when one extra Pb atom is added in the green circle to fill up all the Si-top sites in the bottom half. After further annealing, this structure is very stable with no lower-energy can be found. From the sideview in Fig.6(f), the aligned Pb columns along the *a*-axis are very clear compared to those in Fig.6(d). This structure shows a stripe pattern across the [1$\bar{2}$10] direction as marked by the red and blue vertical lines in Fig.6(e). The bottom half of R30-Si-3top-m11-A-p1 is in compressive strain, while the top half is in tensile strain. The structure in Fig.6(c-d) can be regarded as the same stripe structure in Fig.6(e-f) with a relaxed vacancy defect.

In terms of energetics, at $\theta_{gr}$=0.64, the stripe structure with the vacancy having the $E_f$ of –0.152 eV in Fig.6(c-d) is only slightly more stable than the –0.144 eV of R0-(8×8) structure in Fig.4(c-d) by 8 meV/Pb, although the two intercalated Pb ML arrangements are quite different and rotated by 30 degree. This small energy difference of 8 meV/Pb shows that these two types of Pb ML structures are close in energy and can coexist during intercalation. Indeed, this stripe structure in Fig.6(e-f) is along the Gr zigzag direction and has a periodicity of 2.3 nm across the [1$\bar{2}$10] direction, which agrees well with the direction and periodicity of the stripe phase found in STM experiments[6, 8, 9, 13]. The tensile region relives the large strain built in the R30-Si-3top structure in Fig.2(e-f) where all Si-top sites have been occupied and Pb layer buckles.

In terms of compressibility of the Pb ML near $\theta_{gr}$=0.64 in the (111)-like and stripe structures, we can use the FCC Pb(111) NN distance of 3.51 Å as the reference. In the (8×8) (111)-like structure, the average Pb NN is 3.34 Å, which is a uniform strain of –5% in all the directions in 2D. In the most densely packed R30-Si-3top, the initial NN is the same as Si-Si on SiC(0001) at only 3.08 Å, thus, high strain induces buckling of the Pb ML and it is not as stable as the lower coverage. To relieve the high strain for more stability at lower coverage, instead of the uniform strain as in the (8×8), the stripe phase relieves strain along one direction. R30-Si-3top can be viewed as 15 rows of 5 atoms along [1$\bar{2}$10] direction. A reduced row number to 13 gives an average 15/13*3.08=3.56 Å for Pb NN distance, which is slightly larger than the Pb(111) 3.51 Å and larger than (8×8) 3.34 Å. In this uniaxial strain relief, the Pb NN distance has a large distribution from 3.39 to 3.86 Å along the red and blue vertical lines as seen in Fig.6(e), which corresponding to a strain of –4% for compressive and +10% for tensile region along the strain reliving [1$\bar{2}$10] direction, while the orthogonal direction of [10$\bar{1}$0] still has a compressive strain as high as –12% from the 3.08 Å.

To put the compressed Pb ML intercalated in SiC(0001)/Gr interface in perspective of the other Pb ML systems, it is helpful to compare with the devil's staircase in Si(111)/Pb[22]. For the two generating phases of $(\sqrt{3} \times \sqrt{3})$ and $(\sqrt{7} \times \sqrt{3})$ in the devil's staircase in Si(111)/Pb, the area per Pb atom is 9.58 and 10.64 Å$^2$ for the high and low coverage, respectively. In comparison, the (8×8) phase here also has 9.58 Å$^2$/Pb, close to the densely packed $(\sqrt{3} \times \sqrt{3})$ in the devil's staircase. But in contrast, the stripe phase has 9.44 Å$^2$/Pb, which is about 2% more densely packed overall than the $(\sqrt{3} \times \sqrt{3})$ and 11% more than the $(\sqrt{7} \times \sqrt{3})$ in the devil's staircase. In terms of the uniaxial compressibility, the $(\sqrt{3} \times \sqrt{3})$ and $(\sqrt{7} \times \sqrt{3})$ in the devil's staircase has a lateral –5% compressive strain, while the stripe phase here has –12% in reference to the bulk Pb(111). Then the strain is relieved in the [1$\bar{2}$10] direction with a range from –4% for compressive and +10% for tensile region. Thus, the Pb ML phases intercalated in the confinement of SiC(0001)/Gr interface are denser than those of the devil's staircase in Si(111)/Pb.

## II-c. Electronic analysis of the stripe structure

To better understand the interactions between the intercalated Pb ML and the SiC(0001)/Gr interface in the stripe structure, we have calculated the electron density difference as defined in Eqn.(2),

$$\Delta \rho = \rho_{SiC/Pb/Gr} - \rho_{SiC/Gr} - \rho_{Pb} \qquad (2)$$

Where $\rho_{SiC/Pb/Gr}$, $\rho_{SiC/Gr}$ and $\rho_{Pb}$ are the electron density calculated at the fixed atomic positions with the fully relaxed SiC(0001)/Pb/Gr structure. We plot in Fig.7 at three different isosurface values for the stripe structure. As shown in Fig.7(a-b) for ±0.003 ($e$/Å$^3$), the interaction depicted by the electron density difference upon Pb intercalation is dominated at the interface between the Pb ML and SiC(0001), not between the Pb ML and Gr. The interaction between Pb ML and SiC(0001) is mostly the electron density depletion (cyan) in the Pb ML and the electron density accumulation (yellow) between Pb and Si due to the charge transfer from Pb to Si. The charge transfer from Pb to Si also manifests in the electron density accumulation underneath the top Si along the vertical Si-C bond inside the SiC slab. There is no electron density difference in the bottom SiC layer passivated with H showing the adequate of the slab thickness. With this charge transfer, an electric dipole is built up across the interface between Pb ML and SiC(0001), which further polarizes the Pb electrons and gives electron density accumulation also on the far side of Pb. Noticeably, there is also a thin layer of electron density depletion just above the top layer of Si on SiC(0001). So the electron density accumulation at the SiC(0001)/Pb interface is not only from Pb but also from the Si dangling bonds. In contrast, there is no electron density rearrangement near the Gr overlayer, showing a very weak interaction between the intercalated Pb ML and Gr, even though the Gr(10×10) lattice is expanded by +8.4% to fit the SiC ($5\sqrt{3} \times 5\sqrt{3}$) supercell in the calculation. Such weak interaction agrees with the observation in ARPES[12] for the Pb-intercalated (10×10) moiré that Gr being charge neutral with the Dirac points still near the Fermi energy.

Increasing the isosurface value to $\pm 0.004$ ($e$/Å$^3$) in Fig.7(c-d) shows the difference between the two regions of strains in the stripe structure. In the top half with tensile strain in Fig.7(c), also the right half in Fig.7(d), there are more electron density depletion just above the top Si than those in the bottom half with compressive strain. This shows more electron contribution from the top Si when it is not directly covered by Pb in the tensile region than that in the compressive region. At higher isosurface value of $\pm 0.011$ ($e$/Å$^3$) in Fig.7(e-f), it shows there are more electron density accumulation at the SiC(0001)/Pb interface in the compressive region than the tensile region. All these differences in the electron density rearrangement between these two regions show that the interaction at the compressive region is more ionic than that in the tensile region. In other words, the tensile region has more metallic character in interaction than that in the compressive region, which also reflects in the much less Si-top sites being directly occupied with Pb atoms in the tensile region.

The difference in the compressive and tensile region can also be seen in the PDOS as plotted in Fig.8, where we select one Pb from each region to compare in (a) and one Si from each region to compare in (b). In Fig.8(a), compared to the Pb in the tensile region (blue), the Pb in the compressive region (red) has a larger band width as seen in the low band edge around –10.8 eV, which is lower than the that of the Pb in the tensile region. This difference is due to the higher atomic density and shorter Pb-Pb distance in the compressive than the tensile region. For the comparison of the Si PDOS between the compressive and tensile region, the largest difference is the band gap for the compressive region (red) just above the Fermi level from 0 to 2 eV being filled with extra bands in the tensile region (blue). These different behaviors in PDOS agree with the more metallic interaction in the tensile region than the compressive region as analyzed from the electron density difference in Fig.7.

## III. Discussion

Despite our above work showing the degeneracy between the Pb(111)-like and strain-relieved stripe Pb ML phases intercalated in SiC(0001)/Gr, there are still questions remaining for the observed (10×10) phase, whose structure is still not fully determined, and our Pb(111)-like structure provides a possible but not the definite solution. Part of the reason is still conflicting information obtained from the different type of experiments performed so far. One of the most direct models suggested is based on the STM image of hexagonal phase which shows a period of 2.5 nm consistent with the (10×10) spot observed with diffraction[6, 12]. This was attributed to a coincidence along the Gr direction with (10×10) unit cells matching Pb(111) (7×7) unit cells with coverage $\theta_{gr}$=49/100=0.49 ML with respect to Gr. However, there is a mismatch between (10×10) Gr with SiC(0001), unlike the Gr(13×13) matching well with SiC ($6\sqrt{3} \times 6\sqrt{3}$). To match SiC ($5\sqrt{3} \times 5\sqrt{3}$), the Gr(10×10) needs to be expanded by +8.4%. The experiment performed at higher temperature 945 °C[14] and with a small number of defects has not shown the (10×10) and the intercalated Pb was determined to be in the R30-Si-3top phase with $\theta_{gr}$=0.75 ML. As discussed above this phase can relax to the stripe phase of Fig.6(e-f) with $\theta_{gr}$=0.65 ML to relieve the high strain. Further support that the intercalated Pb is along the SiC unit cell as seen in constant energy momentum distribution spectra that follow the SiC BZ and not the graphene BZ which is rotated 30° from the SiC BZ[12, 14, 16]. Replica cones of primary Dirac cones are seen at positions defined by reciprocal lattice vectors of the SiC and not the graphene lattice.

In addition, new periodic hexagonal arrangements are seen with STM[16] which is attributed to intercalated Pb atoms rearrangements based on the SiC reciprocal lattice that generates domain walls similar to what is seen in the STM images. In the same study a more detailed model was built based on the well-known $(\sqrt{3} \times \sqrt{3})$ and $(\sqrt{7} \times \sqrt{3})$ high coverage phases of Si(111)/Pb that give rise to a family of Devil's Staircase (DS) phases[22] built hierarchically from the two unit cells because of stress. The coverage of the $(\sqrt{3} \times \sqrt{3})$ is 1.107 and the coverage of the $(\sqrt{7} \times \sqrt{3})$ is 1 when now using the bulk Pb(111) as reference for coverage. It is useful to list briefly all the models and their ratio $r$ to the Pb(111) coverage proposed for the intercalated Pb under SiC(0001)/Gr which are based on the experimental information. The coincidence Pb(111) model has the same coverage as Pb(111), so the ratio is 1. The original R30-Si-3top has $r=1.5$, the stripe model of Fig.6(e-f) has $r=1.32$, the model based on the STM image[16] with domain walls has $r=1.24$. Clearly all these models have more compressed Pb layer than the compressed layer of Si(111)/Pb. It has been shown that the formation of the complex DS phases without errors in the large unit cells of each of the DS phases even well below room temperature is related to the unusual collective diffusion of this highly compressed layer[23-28].

Since the intercalated Pb layer under SiC(0001)/Gr is even more compressed for all the models proposed at least for the ones based on the SiC lattice, it is not unreasonable to expect similar collective diffusion. Although confirming this requires more detailed experiments, preliminary evidence includes the ease to intercalate/de-intercalate at the relatively low temperature of 500 °C and the formation of the network of ridges observed even at room temperature of a mildly saturated sub-surface Pb layer. As already discussed, the periodicity of the stripe phase in Fig.6 is along the SiC direction while the (10×10) supercell is along the Gr direction, so the stripe model we propose here is consistent with both the Gr and SiC periodicities discussed in the previous experiments discussed previously.

## IV. Conclusions

In conclusion, using density functional theory calculations, we have explored the phase stability of different coverages of Pb monolayer (ML) structures intercalated in the SiC(0001)/Gr buffer layer interface with the Gr(10×10) supercell supported on the SiC(0001) ($5\sqrt{3} \times 5\sqrt{3}$) supercell. Due to the larger Pb-Pb nearest neighbor distance than the Si-Si distance on SiC(0001), the proposed full Pb ML at the highest coverage occupying all the Si-top sites has a large compressive strain giving rise to the buckling of the Pb ML. After exploring the lower coverage near 0.50 ML with both Si-top site occupied modeling plumbene and the (7×7) structure, we found this structure full relaxes to an amorphous and more energetically stable structure. However, at higher coverage of 0.64 ML, the (8×8) Pb(111)-like structure is quite stable maintaining the Pb(111) lattice after full relaxation and does not relax to an amorphous structure. Since 0.5ML corresponds to the Pb(111) coverage, the Pb-like (8×8) structure has 30% strain above the metallic Pb(111) phase. But a more stable stripe structure has been found by using thermal annealing combined with molecular dynamics, where one half of the (10×10) supercell ML is in compressive strain with Pb occupying the Si-top sites and the other half is in tensile strain with Pb occupying locations off the Si-top sites. This stripe stricture along the Gr zigzag direction (30 degree from the Gr or along the SiC direction) having a periodicity of 2.3 nm across the [1$\bar{2}$10] direction agrees with the STM local observations of small stripe domains in contact with the hexagonal (10×10)-moiré supercells. Analysis with electron density difference and density of states support each other showing the tensile region has a more metallic character than the compressive region, while both are dominated by the charge transfer from the Pb ML to SiC(0001). The small energy difference of 8 meV per Pb between the stripe and (8×8) structure means the two types of structure are almost degenerate and can coexist, explaining the experimental observations.

## V. Methods

Density functional theory[29, 30] (DFT) calculations have been performed with a plane-wave basis set and projector augmented wave method[31], as implemented in the Vienna Ab-initio Simulation Package[32, 33] (VASP). The exchange-correlation (XC) functional of PBE[34] has been used with van der Waals (vdW) interaction included as D3[35, 36]. The SiC(0001) substrate is modeled with four atomic layers of Si and C passivated with H from bottom. A 15 Å vacuum is used above the supported Pb and Gr layers. To compare with different structures that fit into the Gr (10×10) supercell, these structures have been calculated in the SiC(0001) ($5\sqrt{3} \times 5\sqrt{3}$) supercell. We have used a kinetic energy cutoff of 400 eV, $\Gamma$-centered Monkhorst-Pack[37] k-point mesh of (2×2×1) and a Gaussian smearing of 0.05 eV. The ionic positions and unit cell vectors are fully relaxed with the remaining absolute force on each atom being less than $1\times10^{-2}$ eV/Å. Ab initio molecular dynamics (MD) has been used to anneal[20, 21] the structures until no lower-energy structure is found. The Pb ML formation energy ($E_f$) at the SiC(0001)/Gr interface is calculated as

$$E_f = \frac{1}{N}\left(E_{\text{SiC/Pb/Gr}} - E_{\text{SiC/Gr}}\right) - E_{\text{Pb}} \qquad (1)$$

where $N$ is number of Pb atoms in the ML, $E_{\text{SiC/Pb/Gr}}$ is the total energy of the intercalated system, $E_{\text{SiC/Gr}}$ is the total energy of the SiC(0001)/Gr BL without Pb, and $E_{\text{Pb}}$ is the energy of bulk Pb in the face-centered cubic (FCC) structure. With PBE+D3, the relaxed lattice constants of 6H-SiC are 3.085 and 15.139 Å, which agree well with the experimental[38] 3.081 and 15.117 Å. The relaxed lattice constants of graphite are 2.467 and 6.946 Å, which also agree well with the experimental[39] 2.464 and 6.711 Å. The optimized lattice constant for FCC Pb of 4.970 Å is in a good agreement with the experimental[40] 4.950 Å.

Spot profile analysis low energy electron diffraction (SPA-LEED) experiments were carried out in ultra-high vacuum at a base pressure below $1 \times 10^{-10}$ mbar. The growth of a mixture of buffer layer and single layer graphene on 4-H SiC(0001) (Cree Inc) was carried out at high temperatures of about 1200 °C with short 15s flashes performed by an e-beam heater. This process was in-situ monitored by a SPA-LEED spectrometer. Pb was deposited using a flux rate of ∼1/25 monolayer per minute at a low sample temperature of −180 °C (liquid nitrogen). The Pb-covered samples were subsequently annealed using a radiative heater mounted beneath the sample. In this case, temperature was measured with a Re-W (3%–25%) thermocouple.

**Data Availability**: The data that support the findings of this study are available from the corresponding author upon reasonable request.

## Acknowledgements

This work was supported by the U.S. Department of Energy Office of Science, Office of Basic Energy Sciences through the Ames National Laboratory. The Ames National Laboratory is operated for the U.S. Department of Energy by Iowa State University under Contract No. DE-AC02-07CH11358. Some of this research used resources of the National Energy Research Scientific Computing Center (NERSC), a DOE Office of Science User Facility.

**Competing Interests**: The authors declare no competing interests.

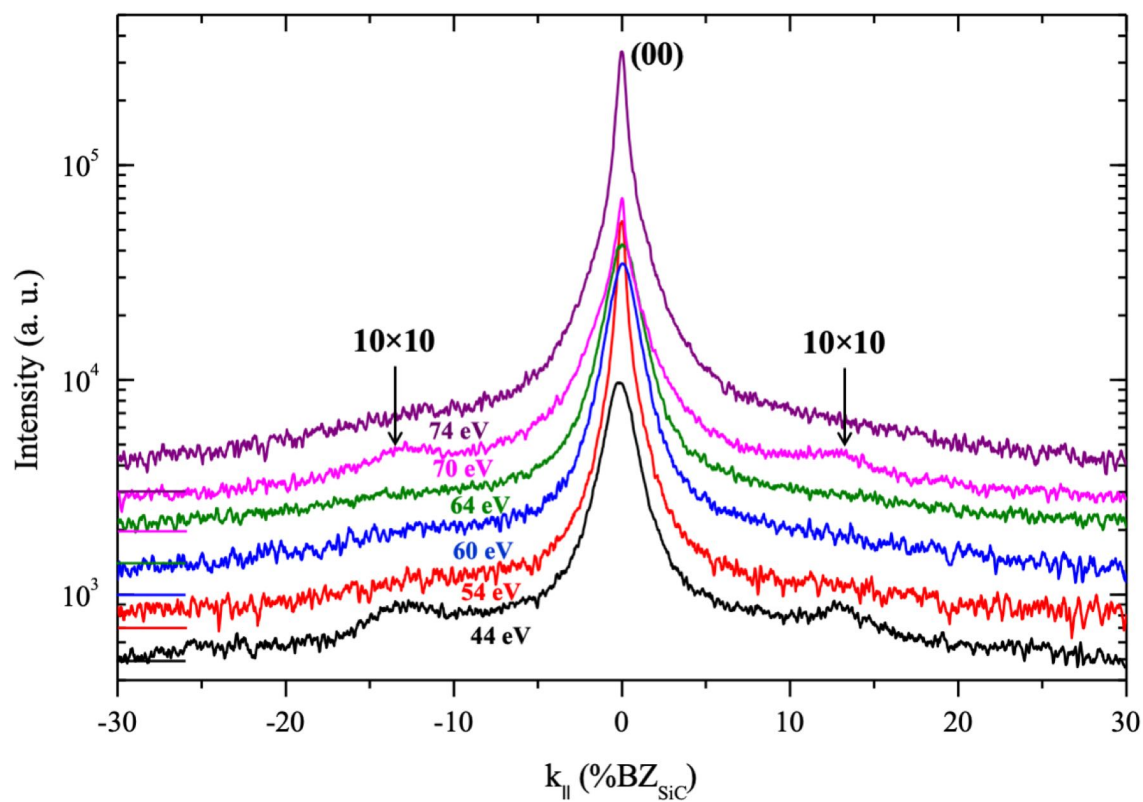

Figure 1. Diffraction profiles obtained with Spot Profile Analysis Low Energy Electron Diffraction (SPA-LEED) close to the 00 spot along the Brillouin zone (BZ) of SiC as a function of energy for intercalating Pb in the epitaxial SiC(0001)/Gr interface. From bottom to top are 44, 54, 60, 64, 70 and 74 eV. The broad peaks in 44 eV (black) and 70 eV (magenta) at $\pm 12.5\%$ degree correspond to the Gr(10×10) supercell. These were obtained after deposition of 13 ML on a surface already intercalated by Pb and gradual annealing from 200 ºC to 450 ºC for 1h at each temperature.

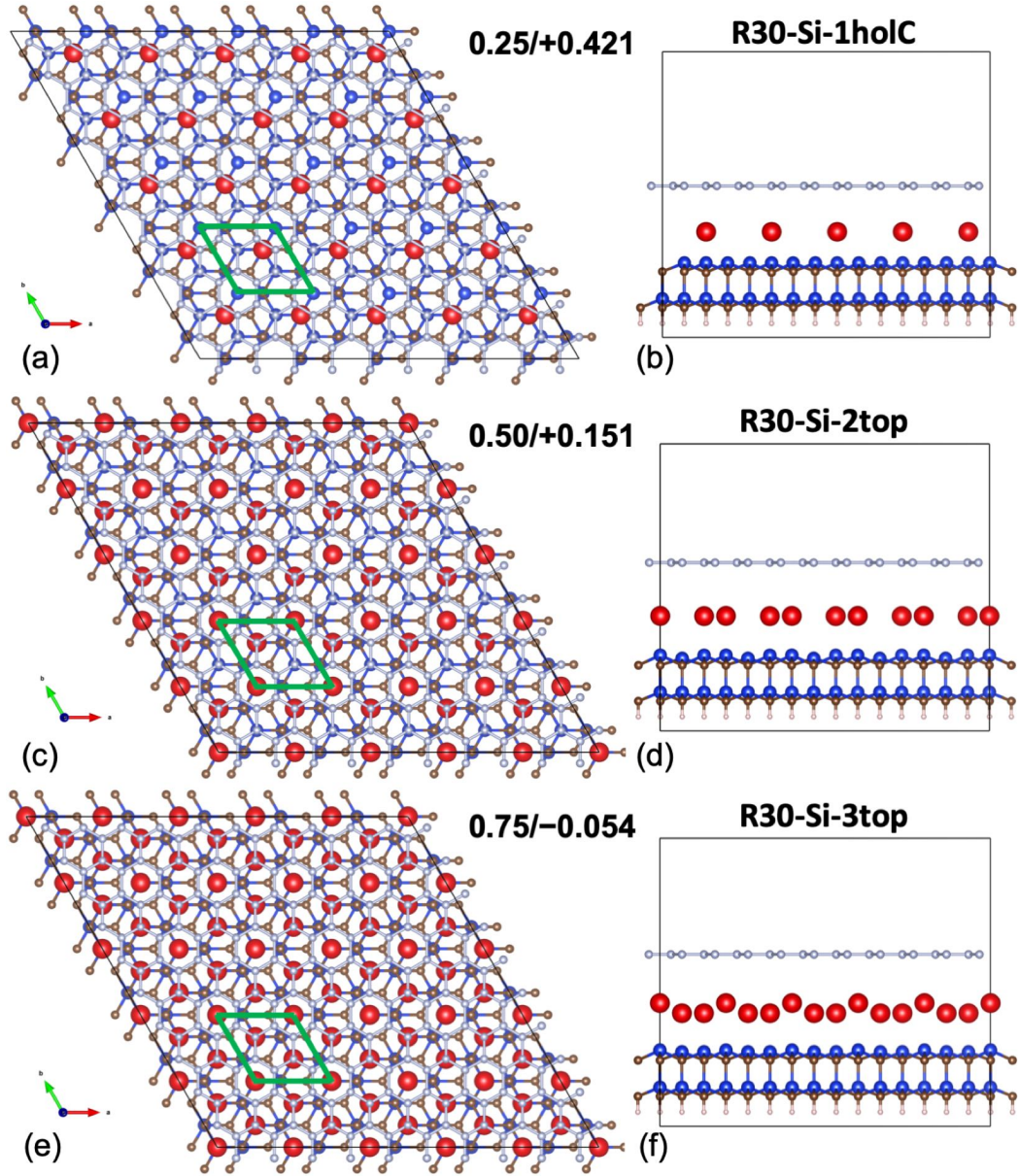

Figure 2. Structures of Pb intercalation inside SiC(0001)/Gr interface at high-symmetry sites with different coverages. The SiC(0001) substrate is modeled by four atomic layers of Si and C passivated with H from bottom. The Si (blue), C (brown), H (pink), Pb (red) and Gr (grey) atoms are shown in different colors. The three structures (a-b) R30-Si-1holC, (c-d) R30-Si-2top and (e-f) R30-Si-3top are shown in both top-view (left) along the $c$-axis and also sideview along the $a$-axis. The two numbers are the coverage with respect to Gr primitive cell ($\theta_{\mathrm{gr}}$) and the formation energy ($E_{\mathrm{f}}$) as defined in Eqn.(1).

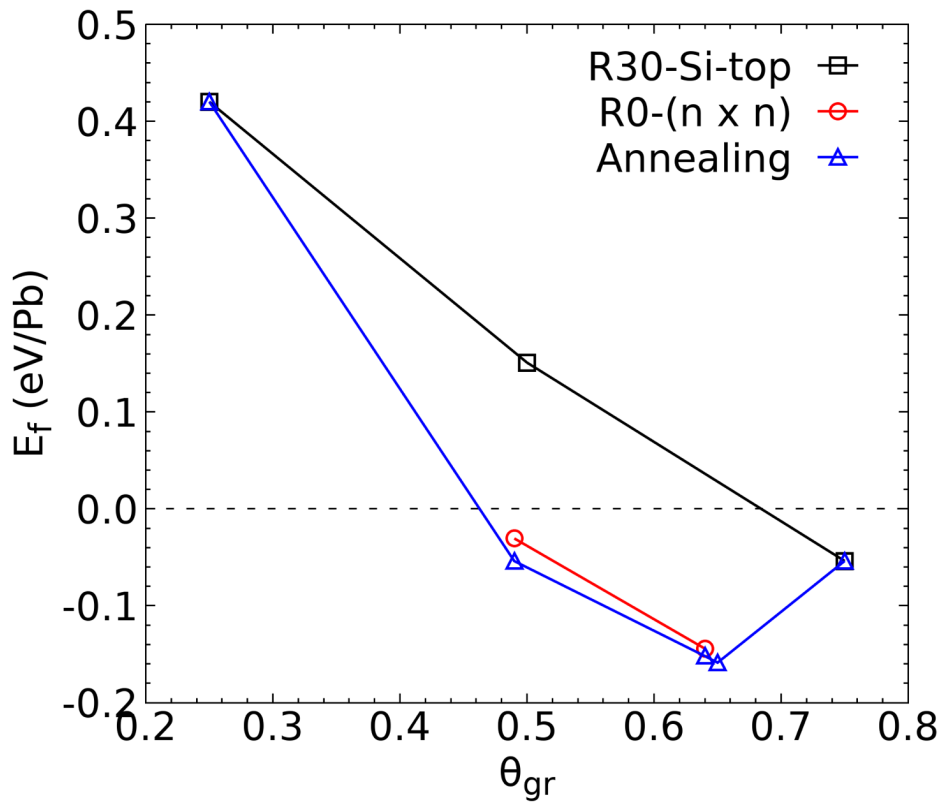

Figure 3. Calculated formation energy ($E_f$) per Pb at different coverages for three types of intercalated Pb monolayer (ML) structures. The R30-Si-top structures are for Pb on high-symmetry sites approaching the highest monolayer coverage with every Si-top site being occupied with Pb and forming the Pb ML in-plane orientation rotated 30 degree from Gr. The R0-(n×n) are for Pb ML with initial Pb(111) orientation 0 degree from Gr with n=7 and 8. Annealing are for the structures obtained after rounds of thermal annealing with molecular dynamics to overcome local energy minimum until no new lower-energy structures are found with the different initial structures from the R30-Si-top structures above. The coverage ($\theta_{gr}$) is with respect to Gr primitive cell and the formation energy ($E_f$) is defined in Eqn.(1).

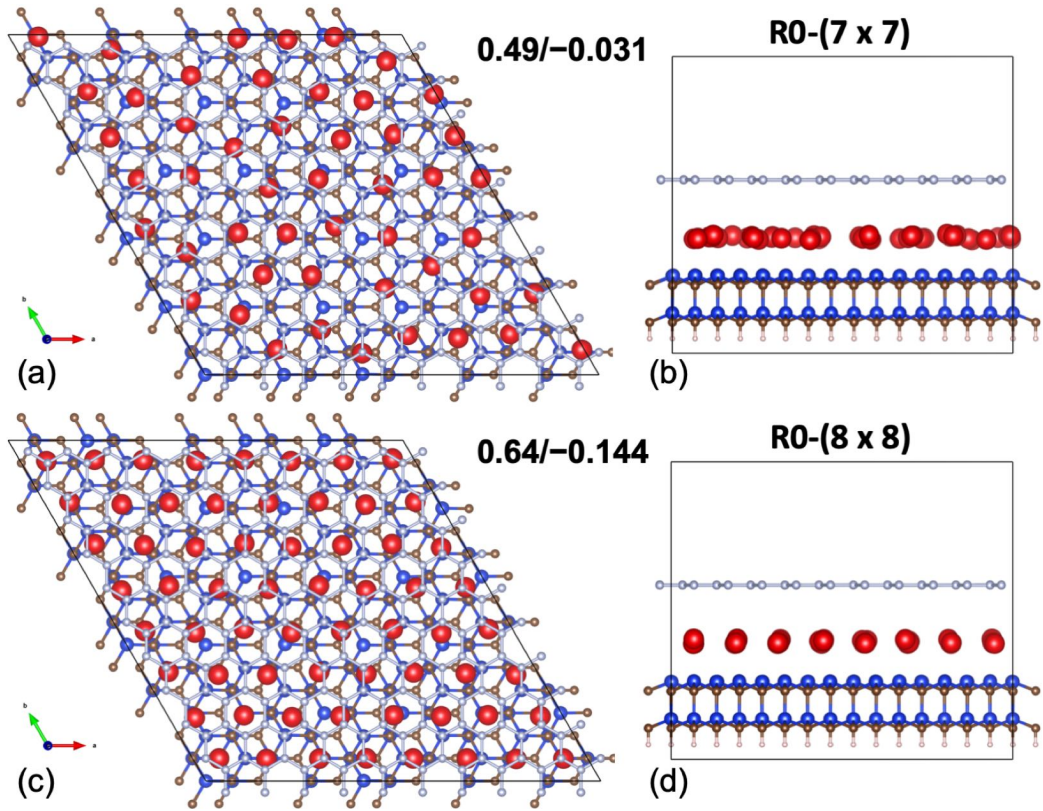

Figure 4. R0-(n×n) structures of Pb intercalation inside SiC(0001)/Gr interface. These are for Pb monolayer (ML) with initial Pb(111) orientation 0 degree from Gr with n=7 and 8. The SiC(0001) substrate is modeled by four atomic layers of Si and C passivated with H from bottom. The Si (blue), C (brown), H (pink), Pb (red) and Gr (grey) atoms are shown in different colors. The two structures (a-b) R0-(7×7) for n=7 and (c-d) R0-(8×8) for n=8 are shown in both top-view (left) along the *c*-axis and also sideview along the *a*-axis. The two numbers are the coverage with respect to Gr primitive cell ($\theta_{gr}$) and the formation energy ($E_f$) as defined in Eqn.(1).

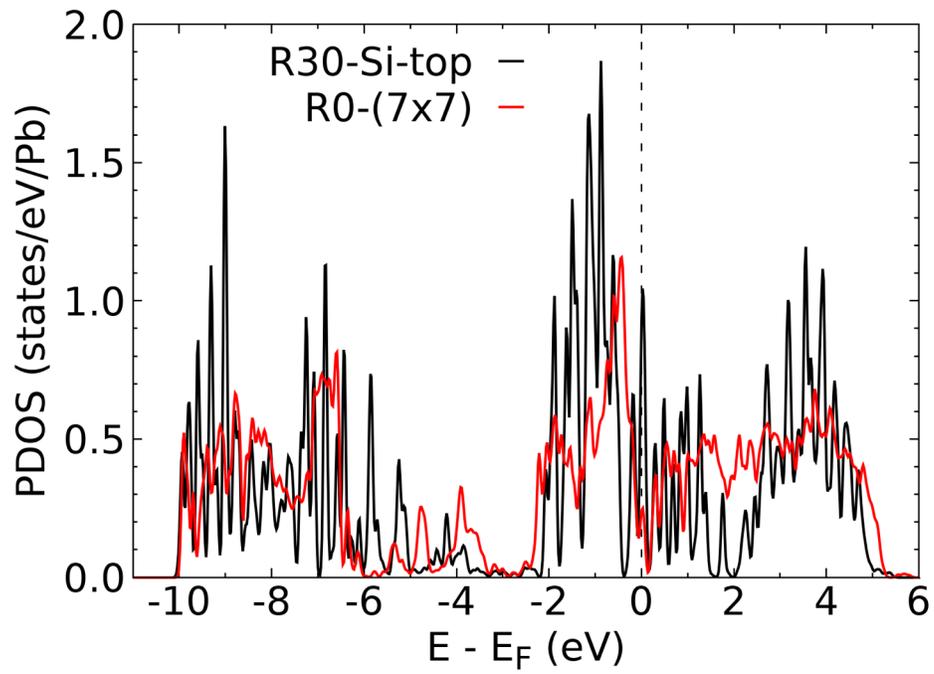

Figure 5. Projected density of states (PDOS) per Pb for two structures. The R30-Si-2top in Fig.2(c-d) (or plumbene) and R0-(7×7) in Fig.4(a-b) are plotted in black and red, respectively.

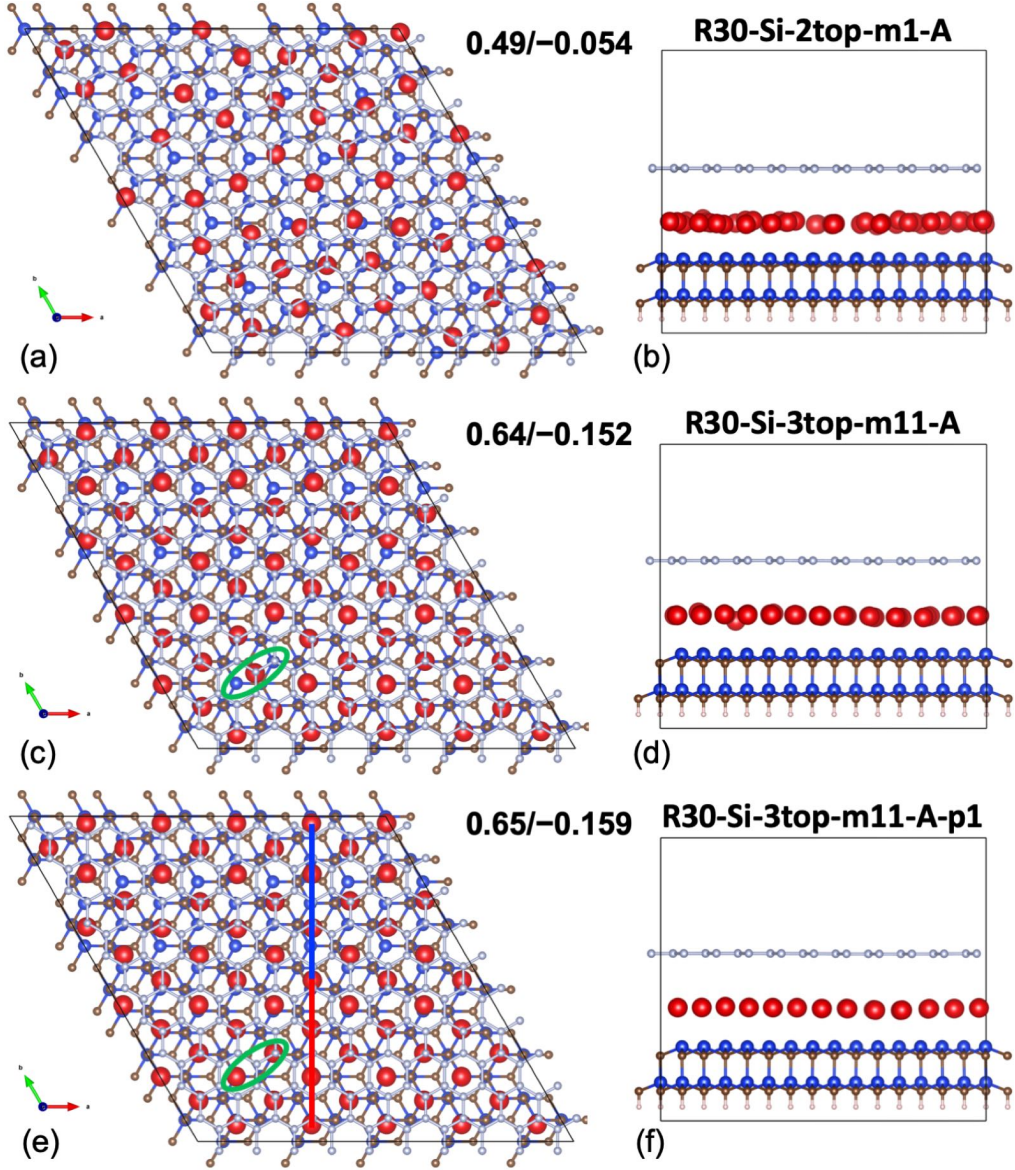

Figure 6. Annealed structures of Pb intercalation inside SiC(0001)/Gr interface at different coverages. The SiC(0001) substrate is modeled by four atomic layers of Si and C passivated with H from bottom. The Si (blue), C (brown), H (pink), Pb (red) and Gr (grey) atoms are shown in different colors. The three structures are shown in both top-view (left) along the $c$-axis and also sideview along the $a$-axis. (a-b) R30-Si-2top-m1-A for annealing started with R30-Si-2top minus 1 Pb atom to make the coverage of 0.49. (c-d) R30-Si-3top-m11-A for annealing started with R30-Si-3top minus 11 Pb atoms to make the coverage of 0.64. (e-f) R30-Si-3top-m11-A-p1 for annealing started with the lowest-energy structure found with R30-Si-3top-m11-A then adding 1 more Pb to make the Si in green circle in (c) fully covered with Pb in. The two numbers are the coverage with respect to Gr primitive cell ($\theta_{gr}$) and the formation energy ($E_f$) as defined in Eqn.(1). In (e), the vertical red and blue lines mark the compressive and tensile strained regions, respectively.

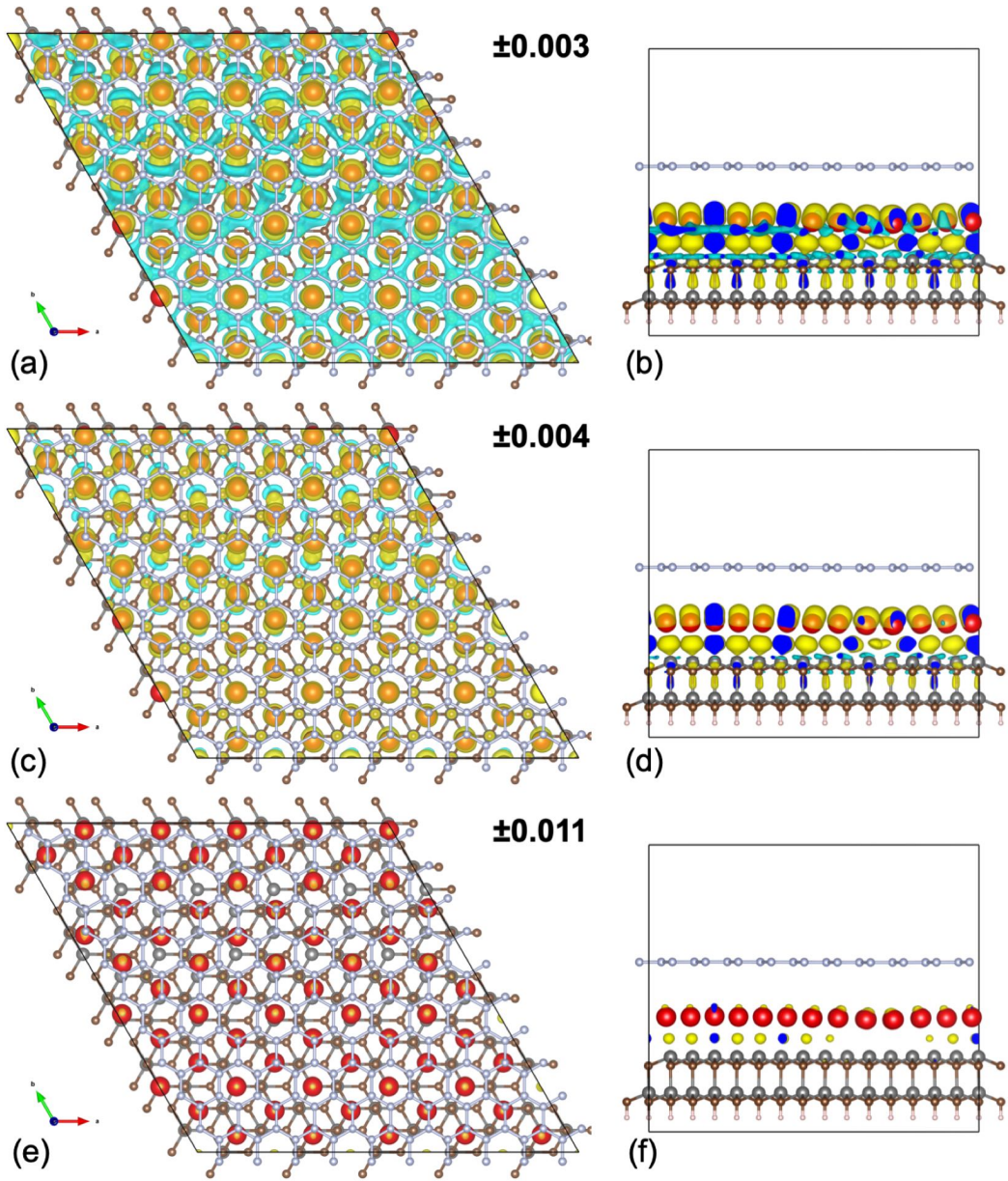

Figure 7. Electron density difference for the stripe structure of R30-Si-3top-m11-A-p1 in Fig.6(e-f). The Si (dark grey), C (brown), H (pink), Pb (red) and Gr (grey) atoms are shown in different colors. The three isosurface values of the electron density difference are shown in both top-view (left) along the $c$-axis and also sideview along the $a$-axis with (a-b) for $\pm 0.003$ ($e$/Å$^3$), (c-d) for $\pm 0.004$ ($e$/Å$^3$) and (e-f) for $\pm 0.011$ ($e$/Å$^3$). The positive and negative electron density change are in yellow and cyan, respectively, on the outside and blue on the inside.

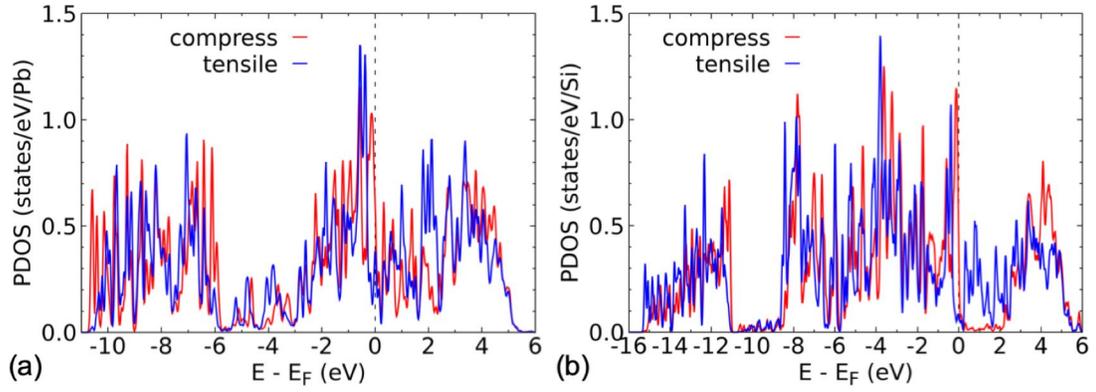

Figure 8. Projected density of states (PDOS) for the stripe structure R30-Si-3top-m11-A-p1 in Fig.6(e-f). One (a) Pb and (b) Si atom in the compressive and tensile region of the stripe structure are selected for comparison.